\documentclass[twocolumn,showpacs,preprintnumbers,amsmath,amssymb]{revtex4}

\usepackage{graphicx}
\usepackage{dcolumn}

\input{epsf}

\newcommand{\bm}[1]{\mbox{\boldmath ${#1}$}}

\newcommand{\scm}[1]{\scriptsize{\mbox {#1}}}
\newcommand{\fnbmk}{\footnotesize{\bmk}}

\newcommand{\bmk}{\bm k}
\newcommand{\ep}{\epsilon}
\newcommand{\om}{\omega}
\newcommand{\Si}{\Sigma}
\newcommand{\si}{\sigma}

\newcommand{\iomn}{i\om_n}

\newcommand{\eg}{{\it e}_{\rm g}}
\newcommand{\tg}{{\it t}_{\rm 2g}}

\newcommand{\eqar}{&=&}

\newcommand{\fr}[2]{\frac{#1}{#2}}

\begin{document}

\title{$\bmk$-dependent spectrum and optical conductivity near metal-insulator transition in multi-orbital Hubbard bands }

\author{Oki Miura}
\author{Takeo Fujiwara}

\affiliation{Department of Applied Physics, University of Tokyo, Bunkyo-ku,
Tokyo 113-8656, Japan}

\begin{abstract}
We apply the dynamical mean field theory
(DMFT)
in the iterative perturbation theory~(IPT)
 to doubly degenerate $\eg$ bands and triply degenerate $\tg$ bands on a simple cubic lattice and calculate the spectrum and optical conductivity in arbitrary electron occupation.~
The spectrum simultaneously shows the effects of multiplet structure and DMFT together with the electron ionization and affinity levels of different electron occupations,~coherent peaks at the Fermi energy  in the metallic phase and a gap at an integer filling of electrons for sufficiently large Coulomb $U$.~
We also calculate the critical value of the Coulomb $U$  for degenerate orbitals.~
\end{abstract}


\maketitle

\section{INTRODUCTION}
 
Many physical properties in strongly correlated systems have been studied 
extensively both experimentally and theoretically.~These are not well described,~however,~by the local density approximation (LDA) based on the density functional theory (DFT).~
LDA+U method
has been developed beyond LDA,~\cite{re:LDA+U2} 
and it works reasonably well for the insulating case near  metal-insulator transition in strongly correlated systems.~
However,~low-energy excitations in metallic phase near metal-insulator transition can not be treated by LDA+U method  
and more sophisticated approaches are needed.

\par
Recently the dynamical mean field theory~(DMFT) has been developed and applied to some model systems,~which led us to a unified picture of both low- and high-energy excitations in the anomalous metallic phase near  metal-insulator transition.~\cite{re:Rev-DMFT} 
DMFT is based on the mapping of many body systems in bulk onto a single impurity problem,~where the off-site Coulomb interaction can be neglected,~subject to a self-consistency condition.~
The combination of DMFT with LDA is applicable to discuss realistic systems.~\cite{re:Ans,re:Lic}~
The combination of DMFT with the GW approximation has been developed to include the off-site Coulomb interaction.~\cite{re:GW}

\par
To solve the mapped single impurity problem within DMFT,~one can use several computational schemes such as the quantum Monte Carlo method~(QMC),~\cite{re:Rozen-QMC,re:Geo-QMC}~the iterative perturbation theory~(IPT),~\cite{re:Kaj-IPT,re:Fuji}~the non-crossing approximation~(NCA)~\cite{re:NCA}~and the exact diagonalization~(ED).~\cite{re:Geo-ED}~
While QMC is not applicable in the low temperature limit,~ED yields only a discrete number of poles for the density of states.~
Moreover,~the exact calculations such as QMC and ED can only be suitable for simple model Hamiltonians.~
To carry out realistic calculations in realistic materials,~
the approximate ones such as IPT or NCA would be more suitable 
due to its CPU-time,~
though IPT is not applicable for cases of large Coulomb interactions and 
NCA can not yield the Fermi liquid behavior at low energies 
and in low temperature limit.~\cite{re:NCA-error}

\par
The aim of the present paper is to apply the generalized DMFT-IPT method for multi-bands on lattices~\cite{re:Fuji} to the doubly degenerate $\eg$ bands and the triply degenerate $\tg$ bands on a simple cubic lattice and discuss the $\bmk$-dependent spectra,~the local Green's function and the optical conductivity near metal-insulator transition.~
In Sec.~\ref{sec:FORM} we will give the general formulation of the dynamical mean field theory and the optical conductivity.~
In Sec.~\ref{sec:RESULT} we will show our numerical results such as 
$\bmk$-dependent spectrum,~the imaginary part of the local Green's function and the optical conductivity for both $n=0.5$ and $n<0.5$ cases.~ 
Finally we summarize our results in Sec.~\ref{sec:CONC}.~

\section{FORMULATION}\label{sec:FORM}
\subsection{Hamiltonian}
We proceed with the Hubbard-type Hamiltonian:
\begin{eqnarray}
H &=& \sum_{jm\si}\ep_d^0 c_{jm\si}^{\dag}c_{jm\si}
  +\sum_{jj'\si mm'}h^{jj'}_{mm'\si}c_{jm\si}^{\dag}c_{j'm'\si}
\nonumber\\
& &+  U \sum_{jmm'}n_{jm\uparrow}n_{jm'\downarrow}
   +  U \sum_{j\si m> m' }n_{jm\si}n_{jm'\si} .
\nonumber\\
\label{eqn:lda-ham1}
\end{eqnarray}
The first and second terms are the single-particle tight-binding Hamiltonian.~ 
Note that index $j$ in the sum of Eq.~(\ref{eqn:lda-ham1})~is running for correlated sites,~$\{m\}$ for orbital indices and $\{\si\}$  for spins.~
$U$ is the on-site Coulomb interaction and the exchange interaction is 
neglected here. 

\subsection{Dynamical mean field theory}

Lattice Green's function and local Green's function are defined as
\begin{eqnarray}
&& G_{m\si m'\si '}(\bmk, \iomn )
\nonumber\\
&&\ \ \ \ =[(\iomn+\mu-\ep_d^0)\bm{1}-h(\bmk)-\Si(\iomn)]^{-1}_{m\si m'\si '},
\label{eqn:Gkw}\\
&& G_{m\si m'\si '}(\iomn)= \fr{1}{V}\int d\bmk~G_{m\si m'\si '}(\bmk,\iomn) ,
\label{eqn:Gw}
\end{eqnarray}
where $h(\bmk)$,~$\mu$ and $\bm{1}$ are the single-particle part of the Hamiltonian in $\bmk$-space,~the chemical potential and the unit matrix.~
Here we neglect the $\bmk$-dependence of the self-energy $\Si$ within the framework of DMFT.~
The chemical potential $\mu$
 is determined to satisfy the Luttinger's theorem.~\cite{re:Mul}~\par

The IPT method was developed by Kajueter and Kotliar for non-degenerate orbital~\cite{re:Kaj-IPT} and generalized by Fujiwara et al. for multi-orbital bands~\cite{re:Fuji} on arbitrary electron occupation.~In the IPT scheme,~we assume the self-energy as
\begin{eqnarray}
\Sigma(i\omega_n) \eqar Un(N_{\scm{deg}}-1)
+\frac{A\Sigma^{(2)}(i\omega_n)}{1-B(i\omega_n)\Sigma^{(2)}(i\omega_n)},
\label{eqn:sig-IPT}   
\end{eqnarray}
where $\Sigma^{(2)}(i\omega_n)$ is the second-order self-energy,~$n_d=\sum_{m\si}n_{jm\si}=nN_{\scm{deg}}$ is the total d-electron occupation number,~$n$ is the occupation number per each orbital and $N_{\scm{deg}}$ is the degree of the degeneracy with respect to spins and orbitals.~
The coefficients $A$ and $B(i\omega_n)$ are determined by requiring the self-energy $\Si (\iomn)$ to be exact in the high-frequency limit~$(\iomn \to \infty)$,~and in the atomic limit~$(U\to \infty)$~\cite{re:Fuji}.~
 $\Sigma^{(2)}(i\omega_n)$ is explicitly proportional to $(N_{\scm{deg}}-1)$ and $U^2$.~\par
The local Green's function is calculated in a form
\begin{eqnarray}
&& G_{mm'}(\iomn) \nonumber\\
&&\ \ \ \ =\fr{1}{V}\sum_{\alpha}\int d\bmk U_{m\alpha}(\bmk)G_{\alpha}(\bmk,\iomn)U^{-1}_{\alpha m'}(\bmk) ,
\label{eqn:Giw-def}\\
&& G_{\alpha}(\bmk,\iomn)= \fr{1}{\iomn+\mu -\ep_d^0-\ep_{\alpha}(\bmk)-\Si (\iomn)} ,
\label{eqn:Gkw-def}
\end{eqnarray}
where the Hamiltonian matrix $h(\bmk)$ is diagonalized to be $\ep_{\alpha}(\bmk)$ by the unitary matrix $U(\bmk)$.~
The $\bmk$-integration is carried out by using a generalized tetrahedron method.~\cite{re:Fuji}~
The $8 \times 243$ $\bmk$-points are used in the $\bmk$-integration within the whole Brillouin zone.~
For a paramagnetic and orbital degenerate system,~the local Green's function becomes diagonal after the $\bmk$-integration.~
The Pad\'{e} approximation is adopted for analytic continuation of the Green's function from the Matsubara frequencies $\iomn$ to the real $\om$-axis.~

\vspace{.5cm}

\subsection{Optical conductivity}

From the linear response theory,~the optical conductivity is given as
\begin{eqnarray}
&& \mbox{Re}\,\si_{\mu\mu'}(\om) =
\sum_{\fnbmk \si}\int d\ep \fr{f(\ep)-f(\ep+\om)}{\om}\nonumber\\
&& \ \ \ \ \times\mbox{Tr}\left[\mbox{Im}\,G(\bmk,\ep)  j_\mu(\bmk)  \mbox{Im}\,G(\bmk,\ep+ \om) j_{\mu '}(\bmk)
 \right]
\label{eqn:opt-definition}
\end{eqnarray}
The matrix element of the current operator is defined as 
\begin{eqnarray}
j^{mm'}_{\mu}(\bmk) \eqar
-e \fr{\partial h^{mm'}(\bmk)}{\partial k_{\mu}}
\:\:\:\:\:\:(\mu=x,y,z).
\label{eqn:current-op}
\end{eqnarray} 

\par

\par
From Eqs.(\ref{eqn:Giw-def})~and (\ref{eqn:opt-definition}),~the optical conductivity is then calculated as 
\begin{eqnarray}
\mbox{Re}\,\si_{\mu\mu'}(\om) \eqar
\int d\ep \fr{f(\ep)-f(\ep+\om)}{\om}
\nonumber\\
&\:\:& \times \sum_{\fnbmk \si}\sum_{\alpha \alpha '}I_{\alpha\alpha '}(\bmk,\ep,\ep+\om)
\nonumber\\
&\:\:& \times  \mbox{Tr}\, \left[ A_{\alpha }(\bmk)  j_\mu(\bmk)  A_{\alpha '}(  \bmk) j_{\mu '}(\bmk)
 \right],
\label{eqn:opt-kansei}
\end{eqnarray}
where $A_{\alpha }(\bmk)=U(\bmk)E_{\alpha}U^{-1}(\bmk)$,~
$I_{\alpha\alpha '}(\bmk,\ep_1,\ep_2)=D_{\alpha}(\bmk,\ep_1)D_{\alpha '}(\bmk,\ep_2)$,~
$E_{\alpha}^{mm'}= \delta_{mm'} \delta_{m\alpha}$,~
and $D_{\alpha}(\bmk,\om)=-\fr{1}{\pi}\mbox{Im}\, G_{\alpha}(\bmk,\om)$.~
The $\bmk$-integration in Eq.~(\ref{eqn:opt-kansei})
~should be carefully carried out.~

\begin{figure*}[]
  \begin{center}
  \resizebox{150mm}{!}{\includegraphics{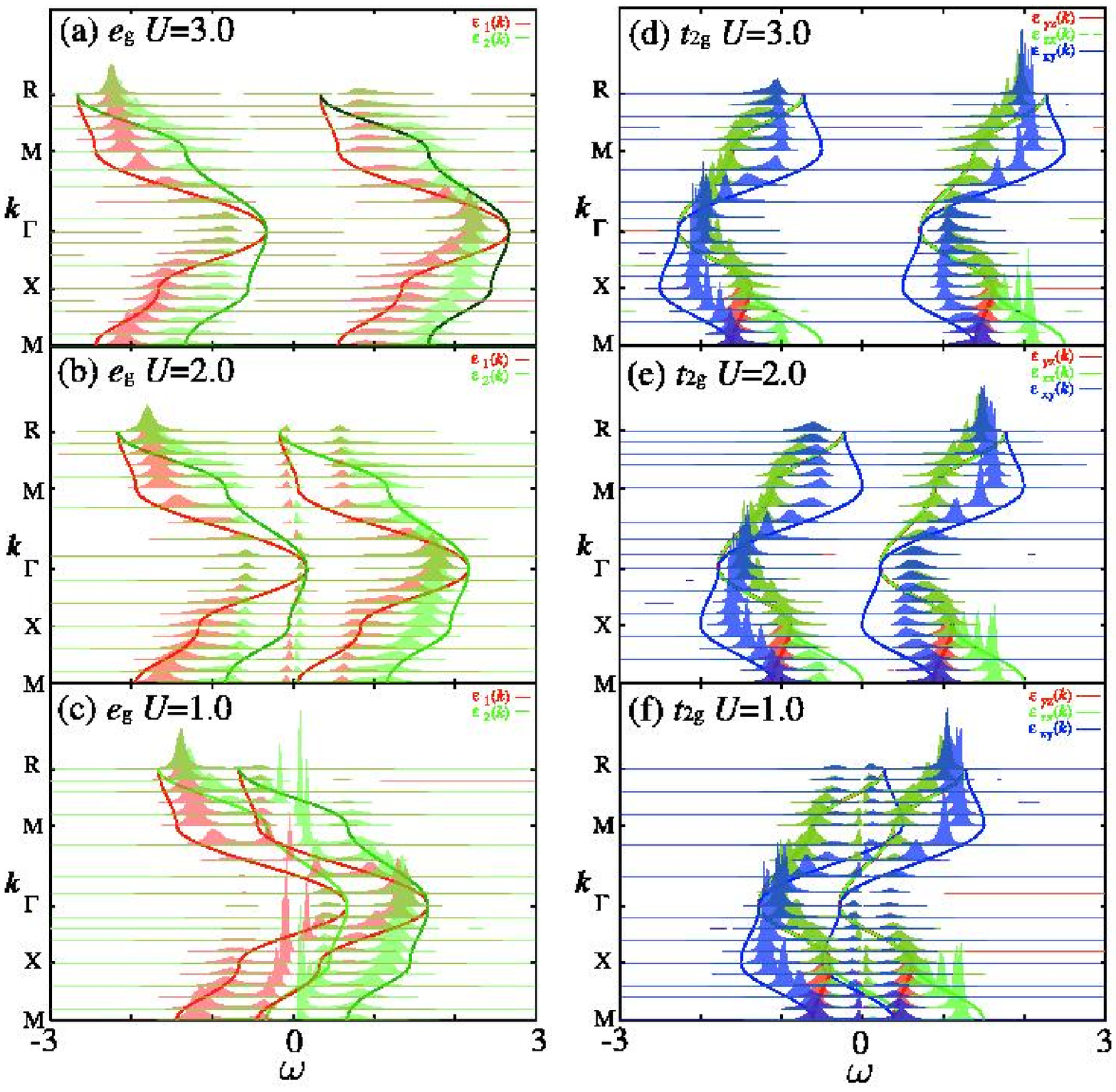}}
\caption{
The imaginary part of the $\bmk$-dependent Green's function $G(\bmk,\om)$~(shaded regions)~of $\eg$ bands and $\tg$ bands on a simple cubic lattice.~
The energy bands $\ep (\bmk)$~shifted by $\pm U/2$~at high-symmetry lines 
are also shown. 
The high-symmetry $\bmk$-points are M$(\fr{\pi}{2},\fr{\pi}{2},0)$,~X$(\fr{\pi}{2},0,0)$,~$\Gamma$$(0,0,0)$,~R$(\fr{\pi}{2},\fr{\pi}{2},\fr{\pi}{2})$.~
(a)~$\eg$ bands,~$U=3.0$.~(b)~$\eg$ bands,~$U=2.0$.~(c)~$\eg$ bands,~$U=1.0$.~
(d)~$\tg$ bands,~$U=3.0$.~(e)~$\tg$ bands,~$U=2.0$.~(f)~$\tg$ bands,~$U=1.0$.~
The inverse temperature $\beta =30$.~
In the $\eg$ bands cases (a)$\sim$(c),~red and green shades represent $G_{11}(\bmk,\om)$ and $G_{22}(\bmk,\om)$ of the diagonalized ${\bmk}$-dependent Green's function matrix,~respectively.~
In the $\tg$ bands cases (d)$\sim$(f),~red, green and blue shades represent $G_{yz,yz}(\bmk,\om)$,~$G_{zx,zx}(\bmk,\om)$~and $G_{xy,xy}(\bmk,\om)$,~respectively.~
}
\label{fig:Gkw} 
  \end{center}
  \end{figure*}

\par
The $\bmk$-integration is able to be carried out simply 
by the generalized tetrahedron method  if the integrand has one pole.~
Since the two-body Green's function $I_{\alpha\alpha '}(\bmk,\ep_1,\ep_2)$ 
has two poles,~we should rewrite it to separate into two partial fractions as 
\begin{eqnarray}
I_{\alpha\alpha '}(\bmk,\ep_1,\ep_2)&=&\fr{1}{2}\mbox{Re}\,\left[G_{\alpha}(\bmk,\ep_1)G_{\alpha '}^*(\bmk,\ep_2)\right.
\nonumber\\
&&-\left.G_{\alpha}(\bmk,\ep_1)G_{\alpha '}(\bmk,\ep_2)
 \right]
\nonumber\\
&=&\fr{1}{2}\mbox{Re}\,\left(
\fr{G_{\alpha}(\bmk,\ep_1)-G_{\alpha '}^*(\bmk,\ep_2)}
{G_{\alpha}^{-1}(\bmk,\ep_1)-\{G_{\alpha '}^*(\bmk,\ep_2)\}^{-1}}
\right.
\nonumber\\
&-&\left.\fr{G_{\alpha}(\bmk,\ep_1)-G_{\alpha '}(\bmk,\ep_2)}
{G_{\alpha}^{-1}(\bmk,\ep_1)-G_{\alpha '}^{-1}(\bmk,\ep_2)}
\right).
\label{eqn:I-henkei}
\end{eqnarray}
In case $\alpha =\alpha'$, the denominator of Eq.~(\ref{eqn:I-henkei}) 
is actually ${\bmk}$-independent and the generalized tetrahedron method 
can be applied. 

\par
When $\om \sim 0$,~we should treat Eq.~(\ref{eqn:I-henkei}) separately 
for $\alpha=\alpha'$ and $\alpha\ne\alpha'$ . 
When $\om \sim 0$ and $\alpha=\alpha'$, Eq.~(\ref{eqn:I-henkei}) 
is then rewritten  as 
\begin{eqnarray}
I_{\alpha\alpha}(\bmk,\ep,\ep)
&=&\fr{1}{2}\mbox{Re}\,\left[|G_{\alpha}(\bmk,\ep)|^2+Z(\ep)\fr{\partial{G_{\alpha}(\bmk,\ep)}}{\partial{\ep}}\right],
\nonumber\\
\label{eqn:I-kansei1}\\
Z(\ep) &=& \left[1-\fr{\partial \Si(\om)}{\partial \om}\Bigg| _{\om \to \ep}\right]^{-1} \ .
\label{eqn:renorm}
\end{eqnarray}
When $\om \sim 0$ and $\alpha \ne \alpha'$, Eq.~(\ref{eqn:I-henkei}) 
should be rewritten as 

\begin{eqnarray}
&& I_{\alpha\alpha '}(\bmk,\ep,\ep)
\nonumber\\
&&\ \ =\fr{1}{2}\mbox{Re}\,
\left[
\{(\ep-\mbox{Re}\,\Si(\ep)-\ep_{\alpha}(\bmk))(\ep-\mbox{Re}\,\Si(\ep)-\ep_{\alpha '}(\bmk))
\right.
\nonumber\\
&&\ \ \ +(\mbox{Im}\,\Si(\ep))^2
+i(\ep_{\alpha}(\bmk)-\ep_{\alpha '}(\bmk))\mbox{Im}\,\Si(\ep)\}^{-1}
\nonumber\\
&&\ \ \ 
\left.
-\fr{G_{\alpha}(\bmk,\ep)G_{\alpha '}(\bmk,\ep)}{\ep_{\alpha}(\bmk)-\ep_{\alpha '}(\bmk)}
\right] \ .
\label{eqn:I-kansei2}
\end{eqnarray}


\begin{figure*}[]
  \begin{center}
  \resizebox{100mm}{!}{\includegraphics{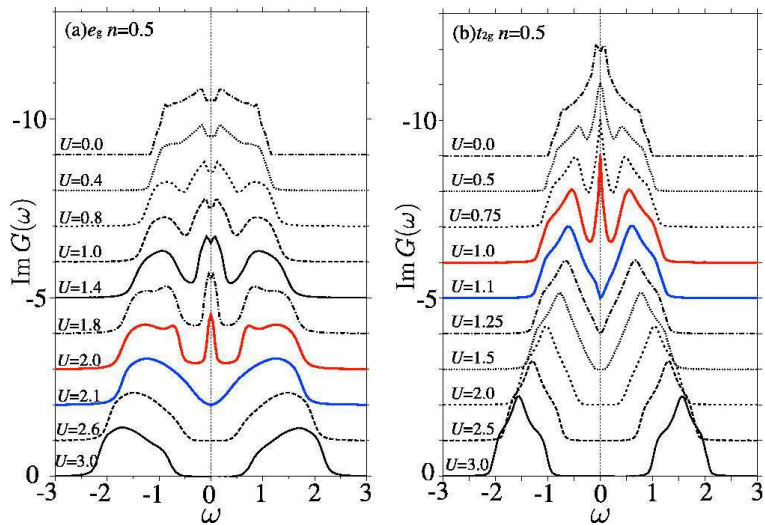}}
  \caption{
The imaginary part of the local Green's functions in a half-filled case $n=0.5$.~
(a)~Doubly degenerate $\eg$ bands of the half band width $D=7/6$.~
(b)~Triply degenerate $\tg$ bands of the half band width $D=1$.~
The inverse temperature $\beta =30$.~
}
\label{fig:GreenCw-n=0.5} 
\end{center}
\end{figure*}

\noindent
In the above two cases, the resultant  $I_{\alpha\alpha '}(\bmk,\ep,\ep)$ 
can be calculated by using the generalized tetrahedron method.

For cases of $\tg$ bands, the ${\bmk}$-dependent Green's function is diagonal 
and this is also  true for $I_{\alpha\alpha '}(\bmk,\ep,\ep)$.
However, for for $\eg$ bands, the ${\bmk}$-dependent Green's function has 
non-zero off-diagonal elements and also  $I_{\alpha\alpha '}(\bmk,\ep,\ep)$ 
has. 
Then the denominators of Eq.~(\ref{eqn:I-henkei}) in case of 
$\alpha \ne \alpha'$ for $\eg$ bands 
are  $\bmk$-dependent 
and, therefore, ${\bmk}$-integration is  not able to be carried out 
by using the generalized tetrahedron method.~ 
For this case, $\bmk$-integration should be carried out 
as the sum of the integrand at each $\om + i\delta$.~

\subsection{The energy bands}

We have applied the above described calculation scheme to the doubly degenerate $\eg$ bands~$(N_{\scm{deg}}=4)$ and the triply degenerate $\tg$ bands~$(N_{\scm{deg}}=6)$.~
We consider Slater-Koster-type tight-binding Hamiltonian on a simple cubic lattice.~
The effective hopping integrals are assumed only between the nearest neighbor pairs and $V_{dd\si}=1/3$,~$V_{dd\pi}=-2/3V_{dd\si}$ and $V_{dd\delta}=1/6V_{dd\si}$.~The relationship among $V_{dd\si}$,~$V_{dd\pi}$ and $V_{dd\delta}$ is the scaling properties of bare two-center integrals in the LMTO method.~Here the half band width of $\eg$ orbitals and $\tg$ orbitals are $D_{\footnotesize{\eg}}=3(V_{dd\si}+V_{dd\delta})=7/6$ and $D_{\footnotesize{\tg}}=2(V_{dd\pi}-V_{dd\delta})=1$,~respectively.~
The off-diagonal elements of $h(\bmk)$ of arbitrary $\bmk$ are non zero for the $\eg$ bands.~
On the other hand,~those for the $\tg$ bands are always zero.~

\section{NUMERICAL RESULTS}\label{sec:RESULT}
\subsection{$\bmk$-dependent spectrum}

The imaginary part of $\bmk$-dependent Green's function $G(\bmk,\om)$ is shown in Fig.~\ref{fig:Gkw}~(shaded regions) with the energy bands $\ep (\bmk)$~(full lines),~for $\beta =30$ and the electron occupation $n=1/2$~(a half-filled case).~
The energy zeroth is set to be at the chemical potential $\mu$.
Energy bands $\ep (\bmk)$~are shifted by $\pm U/2$,~corresponding to lower and upper Hubbard bands.~

\par
Within the Fermi liquid theory,~the $\bmk$-dependent Green's function behaves as 
\begin{eqnarray}
G(\bmk,\om)\simeq Z/\{\om \bm{1}-Zh(\bmk)\}
\label{eqn:Gkw=0}
\end{eqnarray}
near $\om\sim 0$ in a metallic case,~where $Z={(1-\frac{\partial \Sigma}{\partial \omega})}^{-1}$ is a renormalization factor.~
Then the width of the coherent peak at the Fermi energy is scaled by $1/Z$.~
These values become constant within the DMFT,~since the $\bmk$-dependence of the self-energy $\Si$ is $\bmk$-independent.~

\par
In the $\eg$ band case,~peaks appear at the Fermi energy for $U=1.0$ and $U=2.0$,~which shows that the system is in a metallic phase.~
For $U=1.0$,~those peaks strongly depend on $\bmk$ and the spectrum becomes sharper rapidly near the Fermi energy.~
On the contrary,~for $U=2.0$,~those peaks hardly depend on $\bmk$.~
With increasing Coulomb interaction in a metallic phase ($U=2.0$),~coherent peaks appear.~
At $U=3.0$,~the spectrum splits into upper and lower Hubbard bands.~
This results from the combination of electronic structure calculation and DMFT.
The model calculation of DMFT with only assuming the density of states is not able to give the $\bmk$-dependent spectra.~\par
It must be noted that,~in $\tg$ case,~the spectrum shows already the coherent peak hardly depending upon $\bmk$ at $U=1.0$ and it splits into upper and lower Hubbard bands at $U=2.0$.~\par

\begin{figure*}[]
  \begin{center}
  \resizebox{150mm}{!}{\includegraphics{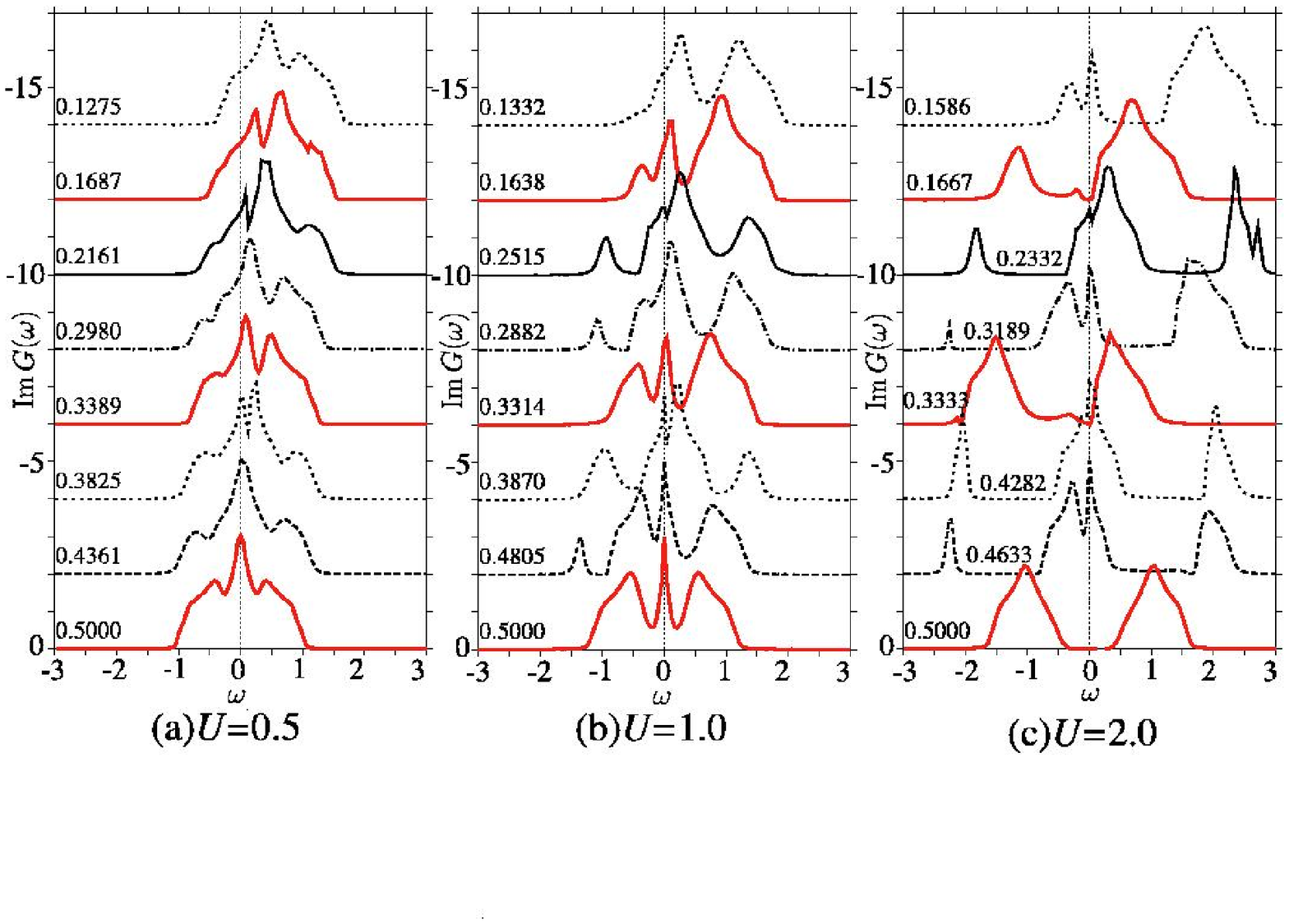}}
  \caption{The electron occupation dependence of the imaginary part of local Green's functions for the triply degenerate $\tg$ bands of the half band width $D=1$~and the inverse temperature $\beta =30$.~
The numbers in the figure denote the electron occupation per each orbital.~
Red line shows the case of the (almost) integer filling of the total d-electron number $n=1/6~(n_d=1),~n=1/3~(n_d=2)$~ and $n=1/2~(n_d=3)$.~ }
\label{fig:GreenCw-U=0.5-1-2}
   \end{center}
\end{figure*}

\subsection{Imaginary part of the local Green's function}
\subsubsection{$n=1/2$~case}

The imaginary part of the local Green's function $G_{mm\si}(\omega)$  
is shown in Fig.~\ref{fig:GreenCw-n=0.5},~for $\beta =30$,~with the electron occupation $n=1/2$~(a half-filled case).~
The positive and negative energy regions,~corresponding to the spectrum of excitation from occupied to unoccupied states (electron ionization spectrum) and that from unoccupied to occupied states (electron affinity spectrum).~

\par
In a half-filled case,~the spectrum consists of the upper and lower Hubbard bands with the electron-hole symmetry.~
At the critical region (here, $U\simeq 2.0$ for $\eg$ bands and $U\simeq 1.0$ for $\tg$ bands) one observes a sharp coherent peak at $\om \simeq 0$.~
In an insulating case,~the self energy shows the singular behavior 
$\Si \sim \fr{1}{\omega}$ near $\omega \sim 0$.~
On the contrary,~In a metallic case,~the self energy behaves as  $\Si -Un(N_{deg}-1)\sim \omega$ near $\omega \sim 0$,~which is typical for the Fermi liquid.~\par
As seen in Fig. \ref{fig:GreenCw-n=0.5},~the critical $U/D$ ratio is different between $\eg$ bands and $\tg$ bands;~$(U/D)_{cr}\simeq1.71$ for $\eg$ bands and $(U/D)_{cr}\simeq1.0$ for $\tg$ bands.~
Figure \ref{fig:GreenCw-n=0.5} also shows that in the insulating case,~the band width is shrinking more at the same $U/D$ ratio for $\tg$ bands than that is for $\eg$ bands.~
This is because that the self-energy is much larger for $\tg$ bands than for $\eg$ bands for the same $U/D$ value due to the degree of the degeneracy with respect to spins and orbitals,~$N_{\scm{deg}}$.~
The second-order self-energy is proportional to $(N_{deg}-1)U^2$ and it contributes $5/3$ times larger for $\tg$ bands than for $\eg$ bands.~
This fact means that the critical $U/D$ ratio of metal-insulator transition 
decreases with increasing degeneracy,~which contradicts to 
the conventional understandings that the critical $U/D$ ratio would be the same as in the s-orbital case,~\cite{Ros} or that the critical $U/D$ ratio 
would increase with it.~\cite{Ham}
These contradiction could originate from the Hamiltonian on a lattice 
or the symmetry type of the orbitals.~\par

\begin{figure*}[]
  \begin{center}
  \resizebox{98mm}{!}{\includegraphics{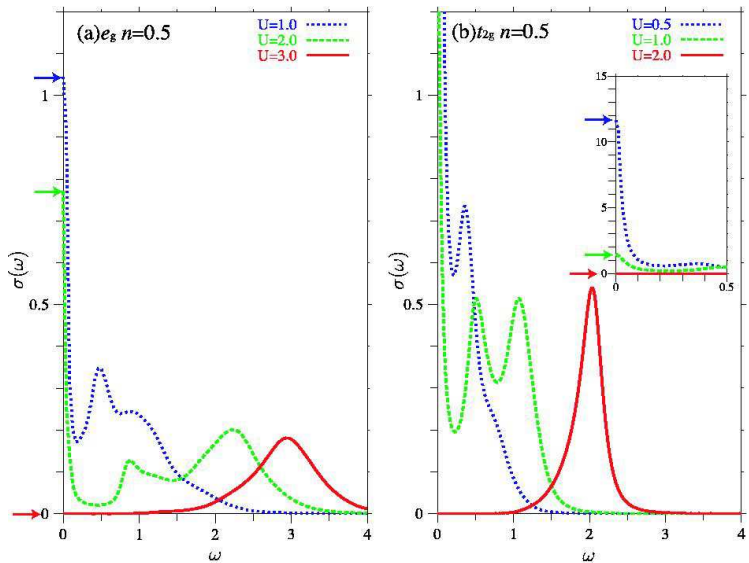}}
  \caption{
The optical conductivity in a half-filled case $n=0.5$.~
(a)~Doubly degenerate $\eg$ bands of the half band width $D=7/6$.~
(b)~Triply degenerate $\tg$ bands of the half band width $D=1$.~
The inverse temperature $\beta =30$.~
The arrows indicate the height of the Drude peak.~
}
\label{fig:opt-cond-n=0.5}
\end{center}
\end{figure*}


\subsubsection{$n<1/2$~case}

Figure \ref{fig:GreenCw-U=0.5-1-2}~shows the electron occupation dependence of the imaginary part of the local Green's functions for $\tg$~ bands with ~$\beta =30$.~
(See Ref.~\onlinecite{re:Fuji} for $\eg$-band.)~
In case of $n=1/6~(n_d=1),~n=1/3~(n_d=2)$~and $n=1/2~(n_d=3)$,~one observes the coherent peak at $\om \simeq 0$ for $U=1.0$,~which is the critical region of  metal-insulator transition.~
We also observes the gap for the same $n$'s at $\om \simeq 0$ for $U=2.0$ and system becomes insulator.~
These occupations correspond to the cases of the integer filling of the total d-electron number.~
At the integer filling case of $n_d=i$~($i$:integer),~the states of $n_d=i$ is completely filled and the states of $n_d=i+1$ is completely empty.~
Then the gap opens between the ionization and affinity spectra.~\par
With small hole doping $\delta$,~in which electron occupation $n_d = i-\delta$,~the spectrum changes drastically at $U=1.0$ and $U=2.0$.~
Since the ground state is the mixture of the states of $n_d=i$ and $n_d=i-1$,~a sharp peak appears below the lower Hubbard band.~
This satellite peak corresponds to the electron ionization of $n_d=i \to i-1$.~
The ionization and affinity levels for both $n_d=i$ and $n_d=i-1$ states appear in the spectrum.~
These satellite structures of all configurations of $n_d=0,1,2,...$ appear but their intensity depends on the occupation number.~
With increasing hole doping $\delta$,~the electron ionization spectrum gradually increases.~
On the contrary,~the electron affinity spectrum loses the intensity.~
When the occupation approaches $n_d=i-1$ from above,~the electron affinity spectrum $n_d=i \to i+1$~disappears.~\par
From above discussion,~we have had success of the spectrum calculation which simultaneously shows the effects of multiplet structure and DMFT such as the electron ionization and affinity levels of different electron occupations,~
coherent peaks at the Fermi energy in the metallic phase
 and a gap at an integer filling of electrons for sufficiently large Coulomb $U$.~

\subsection{Optical conductivity}
\subsubsection{$n=1/2$~case}

Figure \ref{fig:opt-cond-n=0.5} shows the optical conductivity in a half-filled case.~
In a metallic case,~($U=1.0,~2.0$ for $\eg$ bands,~$U=0.5,~1.0$ for $\tg$ bands)~three peaks appear: (1)~the Drude peak at $\om\sim 0$,~(2)~a peak of the width $2D$ at $\om < U$~and (3)~a peak of the width $4D$ at $\om = U$.~
The peak (2) comes from the transition 
from the coherent state to the upper Hubbard band or
that from the lower Hubbard band to the coherent state. 
The peak (3) at $U$~comes from the transition from the lower Hubbard band to the upper Hubbard band.~
The peaks (1) and (2) result from the DMFT which includes the coherent state.~

With increasing Coulomb $U$ in  metallic phase,~contributions of the peaks (1) and (2) decreases.~
In the insulating case,~($U=3.0$ at $\eg$ bands,~$U=2.0,3.0$ at $\tg$ bands)~the peaks (1) and (2) disappear and only that of (3) remains.~
Figure \ref{fig:opt-cond-n=0.5} is consistent with the fact that the critical $U/D$ ratio is different between $\eg$ bands and $\tg$ bands~[$(U/D)_{cr}=1.71$ for $\eg$ bands and $(U/D)_{cr}=1.0$ for $\tg$ bands].~\par

\begin{figure*}[]
  \begin{center}
  \resizebox{150mm}{!}{\includegraphics{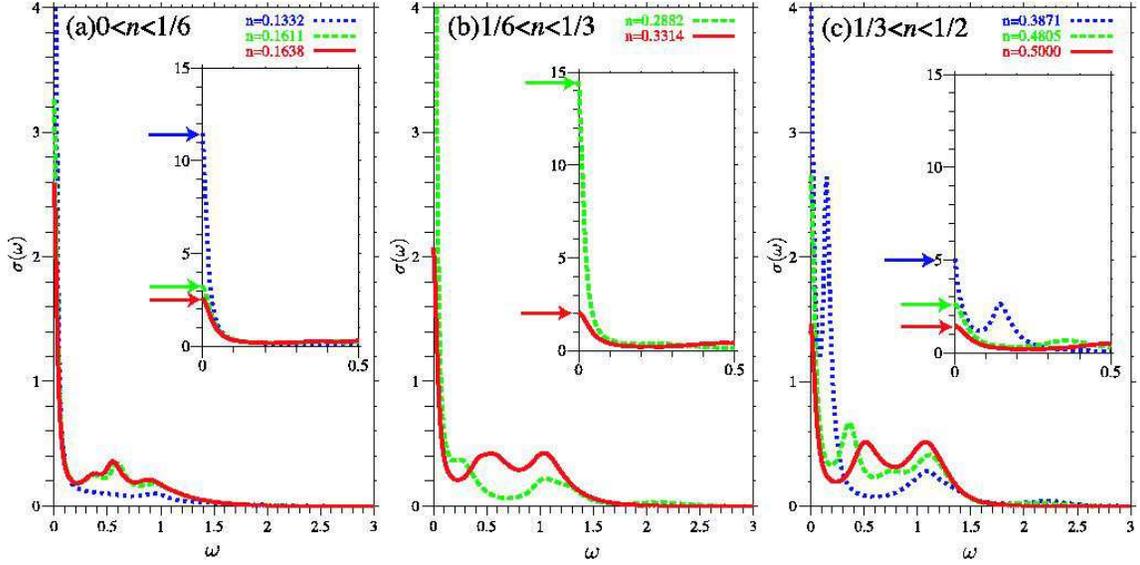}}
  \caption{
The electron occupation dependence of the optical conductivity at $U=1.0$ for the triply degenerate $\tg$ bands of the half band width $D=1$.~
The inverse temperature $\beta =30$.~
The numbers in the figure denote the electron occupation per each orbital.~
(a)~$0 < n \le 1/6$~($0 < n_d \le 1$),~
(b)~$1/6 < n \le 1/3$~($1 < n_d \le 2$),
(c)~$1/3 < n \le 1/2$~($2 < n_d \le 3$).~
The arrows indicate the height of the Drude peak.~
}
\label{fig:opt-cond-t2g-U=1.0} 
  \end{center}
\end{figure*}
\begin{figure*}[]
  \begin{center}
  \resizebox{150mm}{!}{\includegraphics{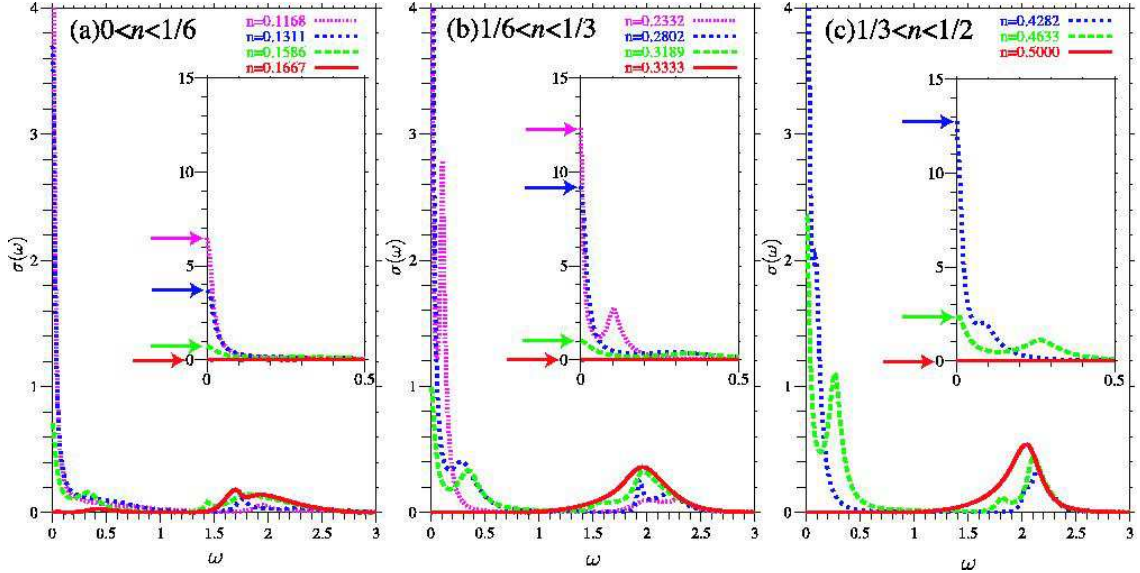}}
  \caption{
The electron occupation dependence of the optical conductivity at $U=2.0$ for the triply degenerate $\tg$ bands of the half band width $D=1$.~
The inverse temperature $\beta =30$.~
The numbers in the figure denote the electron occupation per each orbital.~
(a)~$0 < n \le 1/6$~($0 < n_d \le 1$),~
(b)~$1/6 < n \le 1/3$~($1 < n_d \le 2$),
(c)~$1/3 < n \le 1/2$~($2 < n_d \le 3$).~
The arrows indicate the height of the Drude peak.~
}
\label{fig:opt-cond-t2g-U=2.0} 
\end{center}
\end{figure*}

\subsubsection{$n<1/2$~case}

Figures~\ref{fig:opt-cond-t2g-U=1.0}~and~\ref{fig:opt-cond-t2g-U=2.0} show the optical conductivity in cases of arbitrary electron occupations.~
In case of the integer filling of the total d-electron number,~such as $n=1/6~(n_d=1),~n=1/3~(n_d=2)$~and $n=1/2~(n_d=3)$, 
for $U=2.0$,~the Drude peak disappears  and one observes only a peak at $\om = U$,~which comes from the transition from the lower Hubbard band to the upper Hubbard band.~

With small hole doping $\delta$,~in which electron occupation $n_d = i-\delta$  ($i$=an integer number),~the Drude peak appears and the peak at $\om = U$ shrinks.~One also observes a peak which is separated from Drude peak at $\om \sim 0.4$ for both $U=1.0$ and $U=2.0$.~
With increasing hole doping $\delta$,~this peak shifts toward lower energy and becomes sharper and merges into the Drude peak,~
because it comes from the transition from the lower Hubbard band to 
the coherent state.~
At this range,~the hight of the Drude peak becomes higher and sharper.~ 
Also we can see that the intensity at the lower energy side  of the peak at  
$\omega =U$ decreases simultaneously, and this is because that 
this side of the peak at $\omega=U$ were mainly contributed by 
the transition from the coherent state to the upper Hubbard band. 
When the occupation approaches $n_d=i-1$ from the above,~the Drude peak disappears again for $U=2.0$.~

Optical conductivity successfully shows the ionization and affinity transition and also metal-insulator transition.~This is,~of course,~consistent with the results of the spectrum.~\par

\section{CONCLUSION}\label{sec:CONC}

In conclusion,~we have applied the generalized DMFT-IPT method to doubly degenerate $\eg$ bands and triply degenerate $\tg$ bands.~
The $\bmk$-dependent spectrum shows the resonance behavior near the Fermi energy at smaller $U$ and the coherent peak near metal-insulator transition.~
The spectrum shows that the critical $U/D$ ratio of  metal-insulator transition is different between $\eg$ bands and $\tg$ bands and this originates from the degeneracy~$N_{\scm{deg}}$.~
In the arbitrary occupation case,~the spectrum shows ionization and affinity levels,~coherent peaks at the Fermi energy in a metallic phase and a gap at an integer filling of electrons for sufficiently large Coulomb $U$.~
The optical conductivity gives a spectrum consistent with the above discussions.~


\end{document}